%% file: ANA-EXOT-2018-40-PAPER.tex
\documentclass[cernpreprint,texlive=2016,USenglish,texmf]{atlasdoc}

\usepackage{atlaspackage}
\usepackage{atlasbiblatex}
 
\usepackage{atlasphysics}
\usepackage{atlasjetetmiss}
\usepackage{multirow}
\usepackage{siunitx}
\usepackage{overpic}
\usepackage{tabularx}
\newcolumntype{Y}{>{\centering\arraybackslash}X}

\graphicspath{{logos/}{fig/}}
 
\usepackage{ANA-EXOT-2018-40-PAPER-defs}

\AtlasTitle{Search for dark matter produced in association with a dark Higgs boson
decaying into $W^\pm W^\mp$ or $ZZ$ \\
in fully hadronic final states from $\sqrt{s}=13$ TeV \\
$pp$ collisions recorded with the ATLAS detector
}
\AtlasAbstract{
Several extensions of the Standard Model predict the production of dark matter particles at the LHC.
An uncharted signature of dark matter particles produced in association with $VV=W^\pm W^\mp$ or $ZZ$ pairs from a decay of a dark Higgs boson $s$ is searched for using 139~fb$^{-1}$ of $pp$ collisions recorded by the ATLAS detector at a center-of-mass energy of 13~TeV.
The $s\to V(q\bar q)V(q\bar q)$ decays are reconstructed with a novel technique aimed at resolving the dense topology from boosted $VV$ pairs using jets in the calorimeter and tracking information.
Dark Higgs scenarios with $m_s > 160$~GeV are excluded.
}
\PreprintIdNumber{CERN-EP-2020-172}
 
\author{The ATLAS Collaboration}
 
\AtlasRefCode{EXOT-2018-40}
 
\AtlasDate{December 17, 2020}

\AtlasJournalRef{Phys. Rev. Lett. 126 (2021) 121802}
 
\AtlasDOI{10.1103/PhysRevLett.126.121802}

\hypersetup{pdftitle={ATLAS document},pdfauthor={The ATLAS Collaboration}}
 
\begin{document}
 
\maketitle

Overwhelming astrophysical evidence~\cite{Bertone:2010zza,Feng:2010gw,Porter:2011nv,Bertone:2017adx} suggests the existence of dark matter (DM). DM cannot be accounted for within the Standard Model (SM) and its nature is one of the major questions in physics. 
Several extensions of the SM postulate stable, electrically neutral, weakly interacting massive particles (\wimp)~\cite{Bertone:2017adx} as DM candidates that can potentially be produced in high-energy collisions at the CERN LHC.
Once produced, \wimp would escape detection,
producing an imbalance in the measured transverse momentum\footnote{\AtlasCoordFootnote}, resulting in missing transverse momentum~\ptmiss (with magnitude \met).
A wide class of models probed at the LHC postulate processes where one or more SM particles $X$ are produced recoiling against \wimp, resulting in an ``$X+\met$'' signature. Searches at the LHC have considered $X$ to be a hadronic jet~\cite{EXOT-2016-27,CMS-EXO-16-048}, top or bottom quarks~\cite{SUSY-2016-18,EXOT-2017-16,CMS-EXO-16-049,CMS-EXO-18-010}, a photon~\cite{EXOT-2016-32,CMS-EXO-16-053}, a \W or \Z boson~\cite{EXOT-2016-23,HIGG-2016-28,CMS-EXO-16-037}, or a Higgs boson~\cite{EXOT-2016-25,CMS-EXO-16-055,CMS-EXO-16-050}.
 
This Letter presents a pioneering search for DM using the $X+\met$ signature where $X$ is a hypothetical particle that decays into a vector-boson pair $VV=W^+W^-$ or $ZZ$. This signature was not explored for large \met and resonant $VV$ production with an invariant mass $m_{VV}>160~\GeV$.
The signal region~(SR) requires large \MET from DM particles, and targets the $VV\to q\bar q q\bar q$ decay, which has the largest branching ratio \br.
The background is dominated by vector-boson production in association with jets, referred to as \vjets.
The analysis employs control regions (CRs) requiring either a single muon ($\mu$) or a pair of leptons $\ell^\pm \ell^\mp$ ($\ell = e, \mu$) in the final state to improve background modeling in the SR.

The discovery of a new boson with SM Higgs properties~\cite{HIGG-2012-27,CMS-HIG-12-028,CMS-HIG-12-036} confirmed the mechanism for electroweak symmetry breaking~\cite{Englert:1964et,Higgs:1964ia,Higgs:1964pj,Guralnik:1964eu,Higgs:1966ev,Kibble:1967sv}  and the generation of mass for SM particles. This success motivates a similar mechanism in the dark sector  that contains the DM particle, where \wimp obtains mass via its Yukawa interactions with a dark Higgs boson \s~\cite{Duerr:2017uap}.
Furthermore, \s alleviates the strict constraints from the observed DM relic density~\cite{Aghanim:2018eyx} by  opening up a new annihilation channel into SM particles, when \s, rather than \wimp, is the lightest state in the dark sector.

A two-mediator-based DM model~\cite{Duerr:2016tmh} containing a new \Up gauge symmetry, which yields an additional massive spin-1 vector \Zp boson via the new scalar boson \s, is used for the optimization and interpretation of the search presented in this Letter.
The relevant model parameters are the Majorana DM particle mass \mDM, the \Zp mass \mZp, the dark Higgs mass \ms, and the \Zp couplings \gq to quarks and \gchi to DM particles. The Born-level Feynman diagrams for the process are shown in Figure~\ref{fig:feyn}. The \sDM signal is produced through $q\bar q \to Z' \to s\chi\chi$, requiring an off-shell intermediate state such as a \Zp or $\chi$. The \sWW and \sZZ processes become relevant for $\ms \gtrsim 160~\GeV$ and $\ms \gtrsim 180~\GeV$, respectively~\cite{Heinemeyer:2013tqa}.
The proposed framework shares similarities with previously explored spin-1 simplified DM models~\cite{Buchmueller:2013dya,Harris:2014hga,Buckley:2014fba,Abercrombie:2015wmb,Abdallah:2015ter}, with \s being the only addition and \wimp being a Majorana rather than a Dirac fermion. Within this framework, searches for spin-1 mediators provide complementary sensitivity~\cite{EXOT-2017-32}.

The search is performed using \Lumi~\ifb\ of \pp collisions at $\sqrt{s}=13~\TeV$ recorded with the ATLAS detector~\cite{PERF-2007-01,ATLAS-TDR-2010-19} in 2015--2018. Events in the SR and the single-muon CR were collected by triggering on \met reconstructed from calorimeter information~\cite{TRIG-2016-01} above a threshold that varied from $90$ to $110~\GeV$.
Events in the dilepton CR were recorded using single-lepton triggers with transverse momentum (\pt) thresholds of $24~\GeV$ and higher, depending on the data-taking period, for electrons and muons.
 
SM background processes and the \sDM signal were simulated using Monte Carlo (MC) event generators, except the multijet background, which is found to be negligible using a data-driven method. A detailed simulation of the ATLAS detector~\cite{SOFT-2010-01} based on \GEANT{4}~\cite{Agostinelli:2002hh} was used to simulate the detector response for all MC event samples. Contributions from additional \pp interactions (pileup) were simulated with \PYTHIAV{8.186}~\cite{Sjostrand:2014zea} using the \textsc{NNPDF23} \textsc{LO} parton distribution function (PDF) set~\cite{Ball:2012cx} and corrected to match data. Parton shower simulations with \PYTHIA use the A14 set of tuned parameters~\cite{ATL-PHYS-PUB-2014-021} with the \textsc{NNPDF23} \textsc{LO} PDF set.
 
Signal samples for the $pp \rightarrow \Zp \rightarrow s\chi\chi \rightarrow \V\V\chi\chi \rightarrow \qqbar\qqbar\chi\chi$ process were generated at leading order~(\LO) in QCD with up to one additional parton in the event, using \MGMCatNLOV{2.6.2}~\cite{Alwall:2014hca} interfaced to \PYTHIAV{8.230}, both using the \textsc{NNPDF23} \textsc{LO} \textsc{PDF} set. Samples were generated in the (\mZp, \ms) plane for $\mZp = 0.5, 1, 1.7, 2.5~\TeV$ and in steps of $25~\GeV$ for $160< \ms/\GeV < 360$, with $\mDM=200~\GeV$ to avoid $s\to\chi\chi$ decays. Other parameters were chosen as $\gchi=1.0$, $\gq=0.25$~\cite{Abercrombie:2015wmb,Abdallah:2015ter}, and $\sin\theta=0.01$, where $\theta$ is the mixing angle between SM and dark Higgs bosons~\cite{Duerr:2017uap}, set to a small value~\cite{bib:felix}.
 
The \vjets processes were simulated with \SHERPAV{2.2.1}~\cite{Gleisberg:2008ta}, including mass effects for $b$- and $c$-quarks and using \textsc{NNPDF3.0} PDFs~\cite{Ball:2014uwa}. The perturbative calculations for \vjets were performed at next-to-leading order (\NLO) in QCD for up to two partons and at \LO for up to four partons~\cite{Cascioli:2011va,Denner:2016kdg}, and matched to the parton shower~\cite{Schumann:2007mg} using the ME+PS@NLO prescription~\cite{Hoeche:2012yf}. The \vjets samples are normalized using calculations at next-to-next-to-leading order (NNLO) in QCD~\cite{Anastasiou:2003ds}.
Backgrounds from top quark pair (\ttbar) production and single top quark production were generated at \NLO in QCD with \POWHEGBOX~\cite{Frixione:2007nw,Nason:2004rx,Frixione:2007vw,Alioli:2010xd}~v2 using the \textsc{NNPDF3.0} \NLO PDF set, interfaced to \PYTHIAV{8.230} for parton showering and hadronization. The \ttbar samples are normalized using calculations at NNLO in QCD including next-to-next-to-leading logarithmic corrections for soft-gluon radiation~\cite{Beneke:2011mq,Cacciari:2011hy,Baernreuther:2012ws,Czakon:2012zr,Czakon:2012pz,Czakon:2013goa,Czakon:2011xx}. The single-top-quark processes are normalized to cross sections at \NLO in QCD from Hathor~v2.1~\cite{Aliev:2010zk,Kant:2014oha}.
Diboson ($VV$) samples were simulated with \SHERPAV{2.2.1} at NLO in QCD and normalized using calculations at NNLO in QCD using \textsc{NNPDF3.0} NNLO PDFs.
Backgrounds from associated $VH$ production were generated at \NLO in QCD with \POWHEGBOX interfaced to \PYTHIAV{8.186} using \textsc{NNPDF3.0} NLO PDFs. The $qq \to VH$ and $gg \to VH$ processes were normalized using calculations at NNLO in QCD and at NLO in QCD combined with next-to-leading-logarithmic order corrections, respectively.
 
\begin{figure}[t]
\centering
\includegraphics[width=0.24\textwidth]{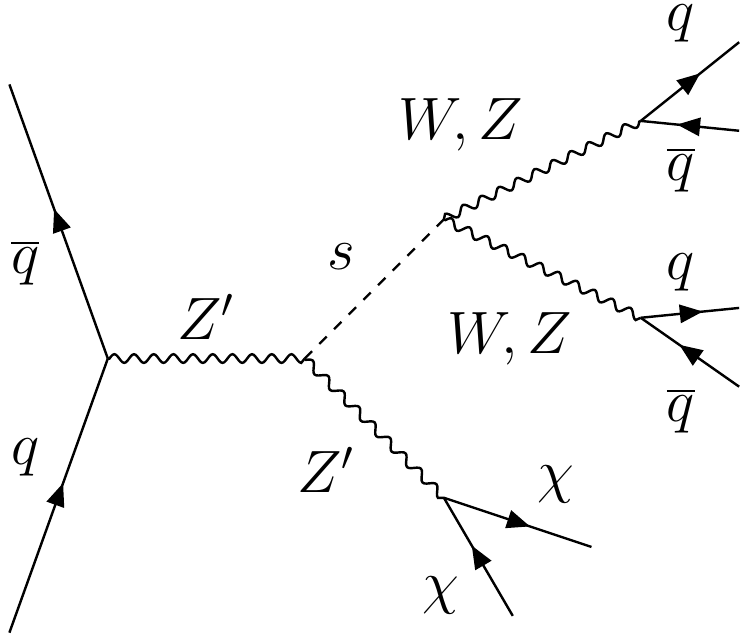}
\quad
\includegraphics[width=0.24\textwidth]{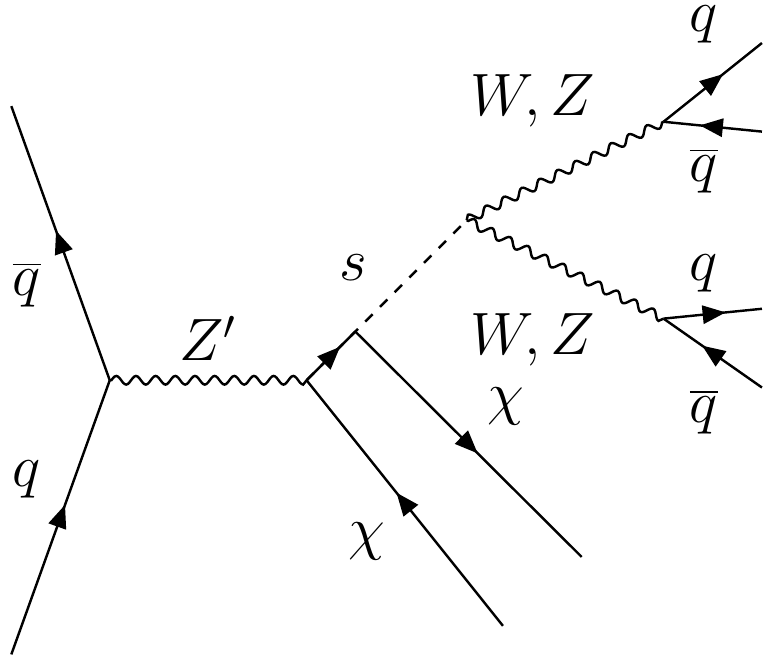}
\caption{
Born-level Feynman diagrams for the $q\bar q \to Z' \to s\chi\chi$, $s\to V(q\bar q) V(q\bar q)$ process. The left diagram typically dominates.
}
\label{fig:feyn}
\end{figure}

At least one \pp collision vertex reconstructed from at least two inner detector (ID) tracks with $\pttrk > 0.5~\GeV$ is required in the event. The vertex with the highest $\sum(\pttrk)^2$ in the event is designated the primary vertex (PV).
The ID tracks must have at least seven hits and satisfy $\pt>0.5~\GeV$ and $|\eta|<2.5$ requirements~\cite{ATL-PHYS-PUB-2015-006,ATL-PHYS-PUB-2015-051}. Their transverse and longitudinal impact parameters relative to the PV must satisfy $|\dzero|<2~\mathrm{mm}$ and $|\zzsth|<3~\mathrm{mm}$, respectively.
 
Muons are reconstructed by matching a track or track segment found in the muon spectrometer to an ID track. Muons must satisfy ``medium'' or ``loose'' requirements~\cite{PERF-2015-10} such that ``medium'' (``loose'') muons must
have $|\eta|<2.5~(2.7)$.
Electrons are reconstructed by matching a cluster of energy in the calorimeter to an ID track. Electron candidates are identified using a likelihood-based method~\cite{EGAM-2018-01} and must satisfy the ``loose'' requirement and have $|\eta|<2.47$.
Electrons and muons must be isolated
according to the track proximity criteria in Ref.~\cite{EXOT-2015-23}.
Hadronic $\tau$-lepton decays are identified by an algorithm based on a boosted decision tree~\cite{PERF-2014-06}.
 
Jets are formed from three-dimensional clusters of calorimeter cells with the \antikt algorithm~\cite{Fastjet,Cacciari:2008gp}.
Small-$R$ jets use a radius parameter $R=0.4$ and are referred to as ``central'' if they satisfy $|\eta|<2.5$ and $\pt>20~\GeV$, and ``forward'' if they fulfill $2.5<|\eta|<4.5$ and $\pt>30~\GeV$. Corrections for pileup~\cite{Cacciari:2008gn} and the energy scale and resolution~\cite{PERF-2016-04} are applied to \smallR jets. In addition, central \smallR jets with $20<\pt/\GeV<60$ and $|\eta|<2.4$ are identified as originating from the PV using associated tracks~\cite{ATLAS-CONF-2014-018}. Small-$R$ jets closer than $\DeltaR = 0.2$ to an $e$, $\mu$, or hadronic $\tau$-decay candidate are rejected.
 
To better reconstruct the challenging multi-prong $s\to V(q\bar q)V(q\bar q)$ decay, the novel track-assisted reclustering (TAR) algorithm~\cite{ATL-PHYS-PUB-2018-012} is used.
This technique improves the resolution of jet substructure observables by considering both tracking and calorimeter information, combined with the flexibility of jet reclustering.
The TAR jets are formed from \smallR jets reclustered into larger jets with $R=0.8$ using trimming parameters optimized for ATLAS~\cite{JETM-2018-02}.
The mass and other substructure observables of TAR jets are reconstructed using ID tracks.
For this, ID tracks are first matched to the \smallR jets that constitute the $R=0.8$ jets. Subsequently, the \pt of tracks matched to a given \smallR jet are rescaled such that their sum equals the \pt of that jet, in order to compensate for the neutral jet components missed by the tracker~\cite{ATL-PHYS-PUB-2018-012}.
The TAR algorithm is estimated to improve the sensitivity of the search by a factor of up to 2.5 in expected median discovery significance compared to the conventional large-$R$ jet approach~\cite{ATLAS-CONF-2016-035}, neglecting systematic uncertainties.
 
In order to suppress contributions from background processes that involve top quarks, which decay almost exclusively to $b$-quarks, a multivariate algorithm is used to identify jets containing $b$-hadrons ($b$-tagging) with an efficiency of 77\%~\cite{FTAG-2018-01}. The algorithm is applied to variable-radius track-jets with $\pt>10~\GeV$ and $|\eta|<2.5$ formed from ID tracks using the \antikt algorithm~\cite{ATL-PHYS-PUB-2017-010} and a \pt-dependent radius parameter.

The \ptmiss vector is computed as the negative vector sum of the transverse momenta of the $e$, $\mu$, and \smallR jet candidates in the event. The transverse momenta not associated with any $e$, $\mu$, or jet candidates are accounted for using ID tracks~\cite{PERF-2016-07}. In addition, an \met significance $\mathcal{S}$ is computed from the expected resolutions for all the objects used in the \met calculation~\cite{ATLAS-CONF-2018-038} and is used to reject multijet background processes.

The signal is characterized by high \met from DM particles, and substantial hadronic activity from $s\to V(q\bar q)V(q\bar q)$ decays that results in an invariant mass consistent with \ms.
Thus, the SR requires $\met>200~\GeV$, no isolated $e$ or $\mu$, no $\tau$ lepton decays, and two or more \smallR jets.
Events in the SR are rejected if a ``loose'' electron or muon with $\pt>7~\GeV$ is present.
In addition, events in the SR and CRs are not considered if they contain hadronic $\tau$-decay candidates with $\pt>20~\GeV$ within $|\eta|<2.5$. The smallest azimuthal angle between the \met and any of the three highest-\pt (leading) \smallR jets is required to be at least $\pi/9$ in order to reduce the multijet background arising from mismeasured jet momenta. This background is further suppressed by requiring $\mathcal{S}>15$.
 
The \ttbar and diboson processes contribute $1\%$--$7\%$ and $2\%$--$8\%$ of the background in the SR, respectively, while the dominant SM \znunujets and $W(\ell\nu)\!+\!{\mathrm{jets}}$ processes contribute $59\%$--$73\%$ and $15\%$--$32\%$, respectively, depending on the topology. The modeling of \vjets is improved using two CRs: the single-muon CR (\CRone) enriched in \wjets, and the two-lepton CR (\CRtwo) enriched in \zjets. The \onemucr follows the same selection as the SR, except that events must contain exactly one ``medium'' muon with $\pt>27~\GeV$ and no ``loose'' electrons with $\pt>7~\GeV$. Events in the \twolepcr are selected using the same requirements as the SR, except that events must contain exactly two ``loose'' electrons or two oppositely charged ``medium'' muons, and satisfy $\mathcal{S}<15$. The leading lepton must fulfill $\pt>27(25)~\GeV$ for electrons (muons), while for the subleading one $\pt>7~\GeV$ is required.
The dilepton system is required to be consistent with an energetic $Z$ boson, i.e.\ $\pt^{\ell\ell}>200~\GeV$ and $83<m_{\ell\ell}/\GeV<99$.
 
In order to optimize the sensitivity over a broad $VV$-pair momentum range, two selection categories, merged and intermediate, are defined.
For large $s$ momenta, the dark Higgs boson's decay products become collimated and are reconstructed inside a single TAR jet. These topologies are targeted in the \merged category, defined as containing at least one TAR jet with $\ptTAR>300~\GeV$, and mass \mtar between $100~\GeV$ and $400~\GeV$. TAR jet substructure variables are employed to discriminate between the four-prong topology of $s\to V(q\bar q)V(q\bar q)$ decays and backgrounds with lower multiplicities. This is done using combinations of $N$-subjettiness~\cite{Thaler:2010tr} variables $\tau_{N}$ by requiring $0<\tau_{4}/\tau_{2}<0.3$ and $0<\tau_{4}/\tau_{3}<0.6$, which were also experimentally studied in Ref.~\cite{Khachatryan:2015bma}. The $s$-candidate mass is identified with \mtar.
The \merged category dominates the sensitivity, and the product of acceptance and selection efficiency for $\sigma(pp \rightarrow s \chi\chi)\times\mathcal{B}(\sVV)$ lies around 1\%.
 
Moderate $s$-candidate momenta result in less-collimated decay products, which may not be captured by the nominal TAR jet. In such cases, events failing the \merged-category requirements are considered in the \intermediate category, where the $s$ candidate is reconstructed from a TAR jet with $\mtar>60~\GeV$ that is supplemented by up to two additional \smallR jets within $\DeltaR=2.5$ of the TAR jet. If the mass of the TAR jet is compatible with $m_W$, i.e.\ $60<\mtar/\GeV<100$, the TAR jet is supplemented with the two \smallR jets whose combined invariant mass is closest to $m_W$.
If $\mtar>100~\GeV$, it is assumed that only one prong of the $s$ decay was not reconstructed within the TAR jet, and thus it is supplemented with exactly one \smallR jet.
The $s$-candidate mass is required to lie between $100~\GeV$ and $400~\GeV$.
The product of acceptance and selection efficiency for $\sigma(pp \rightarrow s \chi\chi)\times\mathcal{B}(\sVV)$ ranges between 10\% and 20\%.

To account for changes in the background composition and benefit from increased signal sensitivity with higher \met, events in the \merged category are further classified into ranges in $\met/\GeV$: $[300,500]$ and $[500,\infty)$. The range $[200,\infty)$ is used in the \intermediate category. The same ranges are defined consistently in the \onemucr and the \twolepcr. To ensure kinematic similarity to the \ptmiss arising from the \vjets in the SR, $\ptmissnomu = \ptmiss+\boldsymbol{p}_\text{T}^\mu$, which corresponds to the \pt carried by the $W$ boson, is used in the \onemucr. Similarly, $\boldsymbol{p}_\text{T}^{\ell\ell}$ in the \twolepcr corresponds to \ptmiss in the SR.
 
The DM signal is extracted via a simultaneous maximum-likelihood fit~\cite{Moneta:2010pm,Verkerke:2003ir_} of signal and background simulations to the binned \s-candidate mass distributions in the SR and to total yields in the CR categories.
The normalizations of \wjets and \zjets processes are free parameters in the fit and are constrained by the total event yields, summed over \met-bin and category, in the \onemucr and  \twolepcr.
Experimental uncertainties related to the calibration of the scale and resolution of the jet energy~\cite{PERF-2016-04} as well as to tracking efficiencies~\cite{ATL-PHYS-PUB-2015-051} affect the reconstruction of \ms using TAR jets. Other leading experimental systematic uncertainties arise from the finite number of MC events and the calibration of the lepton identification efficiencies~\cite{PERF-2015-10,EGAM-2018-01}. Dominant theoretical systematic uncertainties originate from the modeling of the signal and the \wjets and \zjets background processes.
These encompass uncertainties from the choice of PDFs and factorization and normalization scales.
In addition, to estimate the uncertainty from the choice of matrix element and parton shower generator for \wjets and \zjets, alternative MC samples generated with \MGMCatNLOV{2.6.2} at LO in QCD with up to four parton emissions using the \textsc{NNPDF23} \textsc{LO} \textsc{PDF} set and interfaced to \PYTHIAV{8.230} using a merging scale of $Q_\text{cut} = 30~\GeV$ are considered.
All other systematic uncertainties are estimated similarly to Ref.~\cite{EXOT-2016-25}, except for the $\ttbar$ normalization, for which theoretical uncertainties~\cite{Czakon:2013goa} are considered.
The systematic uncertainties, parameterized as nuisance parameters with Gaussian or log-normal prior probabilities, are profiled and used to constrain the template shapes and the normalizations varied in the fit~\cite{Aad:2014xzb}.
Dominant uncertainties after the fit to Asimov data for three representative dark Higgs scenarios are quantified in Table~\ref{tab:systs}.

\begin{table*}
\caption{
Dominant sources of uncertainty for three dark Higgs scenarios after the fit to Asimov data generated from the expected values of the maximum-likelihood estimators including predicted signals with $\mZp = 1~\TeV$ and \ms of (a)~160~\GeV, (b)~235~\GeV, and (c)~310~\GeV.
The uncertainty in the fitted signal yield relative to the theory prediction is presented.
Total is the quadrature sum of statistical and total systematic uncertainties, which consider correlations.
}
\centering
\label{tab:systs}
\begin{tabularx}{0.45\textwidth}{l*{4}{Y}}
\toprule
\multirow{2}{*}{Source of uncertainty} & \multicolumn{3}{c}{Uncertainty [\%]} \\
& (a) & (b) & (c) \\
\midrule
Signal modeling & 11 & 10 & 10 \\
\wjets modeling & 9 & 21 & 14 \\
\zjets modeling & 7 & 12 & 13 \\
MC statistics & 11 & 14 & 23 \\
Jet energy scale & 8 & 17 & 24 \\
Jet energy resolution & 11 & 18 & 15 \\
Lepton reconstruction & 8 & 9 & 5 \\
Track reconstruction & 6 & 7 & 5 \\
\midrule
Systematic uncertainty & 30 & 42 & 55 \\
Statistical uncertainty & 16 & 25 & 50 \\
Total uncertainty & 34 & 49 & 74 \\
\bottomrule
\end{tabularx}
\end{table*}
 
A first fit to the SM backgrounds is performed using only data from the CRs. The observed and fitted yields in the CR categories obtained after this fit are shown in Figure~\ref{fig:yields-body}. Also shown are the background yields predicted in the SR when using the observed parameter values from the CR-only fit. The fit reduces the MC-predicted \vjets contribution in the merged category.
The overall yields in the CRs and the SR are found to be well described by SM simulations.
The normalization and the $\pt^V$ dependence of both \wjets and \zjets are consistent within uncertainties with SM predictions in the SR and CRs.
Figure~\ref{fig:ms-body} shows the mass distributions $m_{VV}$ of the \s candidate in two representative SR categories and the two corresponding categories in the \onemucr, obtained after a simultaneous fit to the SR and the CRs under the hypothesis that only SM predictions are present. The data distributions agree well with MC simulations in the CRs, indicating that \vjets background processes reconstructed with the novel TAR algorithm are well modeled. The observed results in the SR indicate that the data are in general well described by SM predictions.
A mild excess around $m_{VV} = 160~\GeV$ is observed, yielding a $2.3\sigma$ local significance and $1.3\sigma$ global significance when considering nine independent $\ms$ hypotheses. The excess in the intermediate region is narrower than the experimental resolution for \ms.

\begin{figure}
\centering
\includegraphics[width=0.7\textwidth]{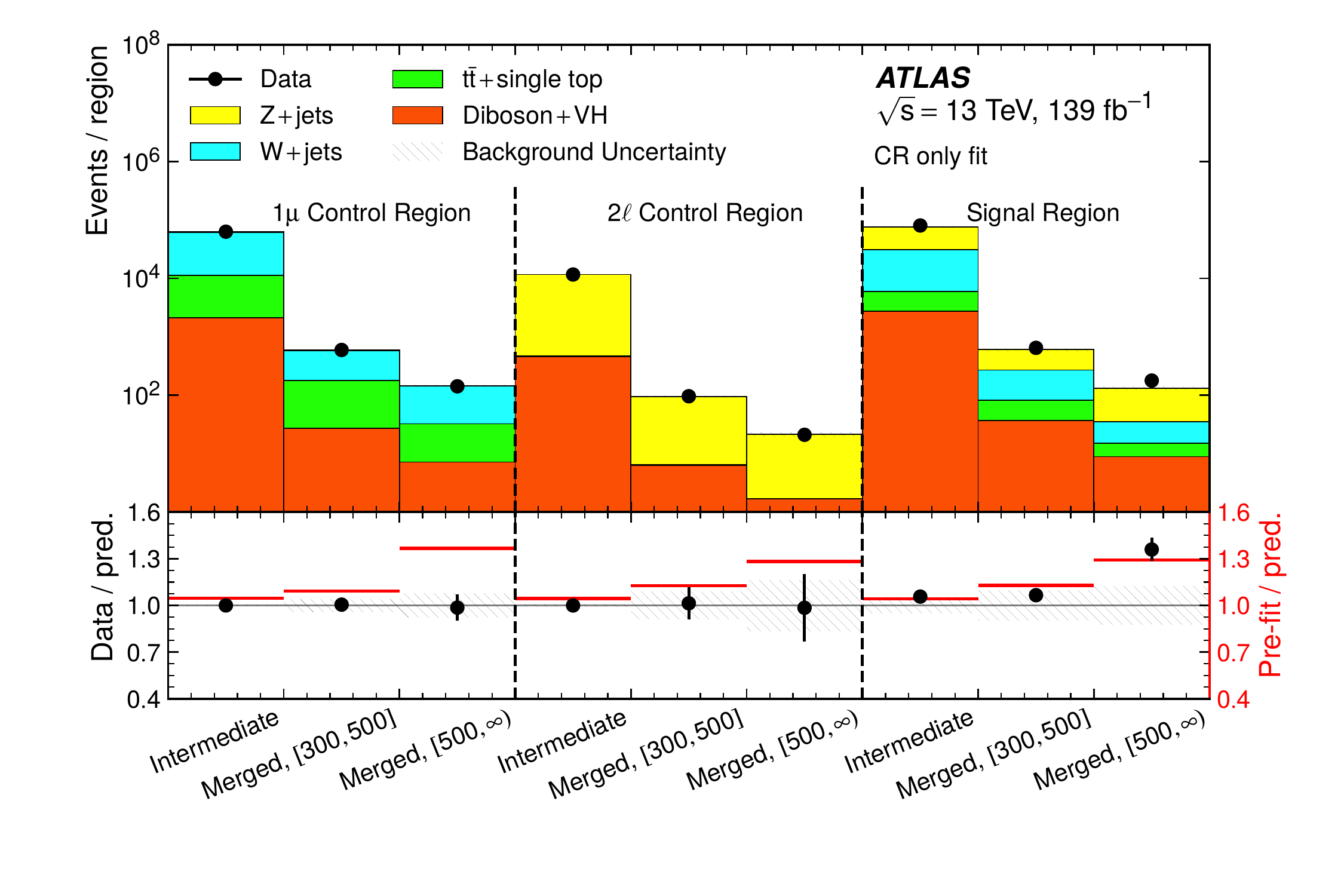}
\caption{
Data overlaid on SM background post-fit yields stacked in each SR and CR category and \met bin with the maximum-likelihood estimators set to the conditional values of the CR-only fit, and propagated to SR and CRs.
The ratio of the data to SM expectations after the CR-only fit is shown in the lower panel, along with the red line representing the ratio of the pre-fit to the post-fit background prediction.
Pre-fit uncertainties cover differences between the data and pre-fit background prediction.
}
\label{fig:yields-body}
\end{figure}

\begin{figure}
\centering
\subfloat[]{\includegraphics[width=0.49\textwidth]{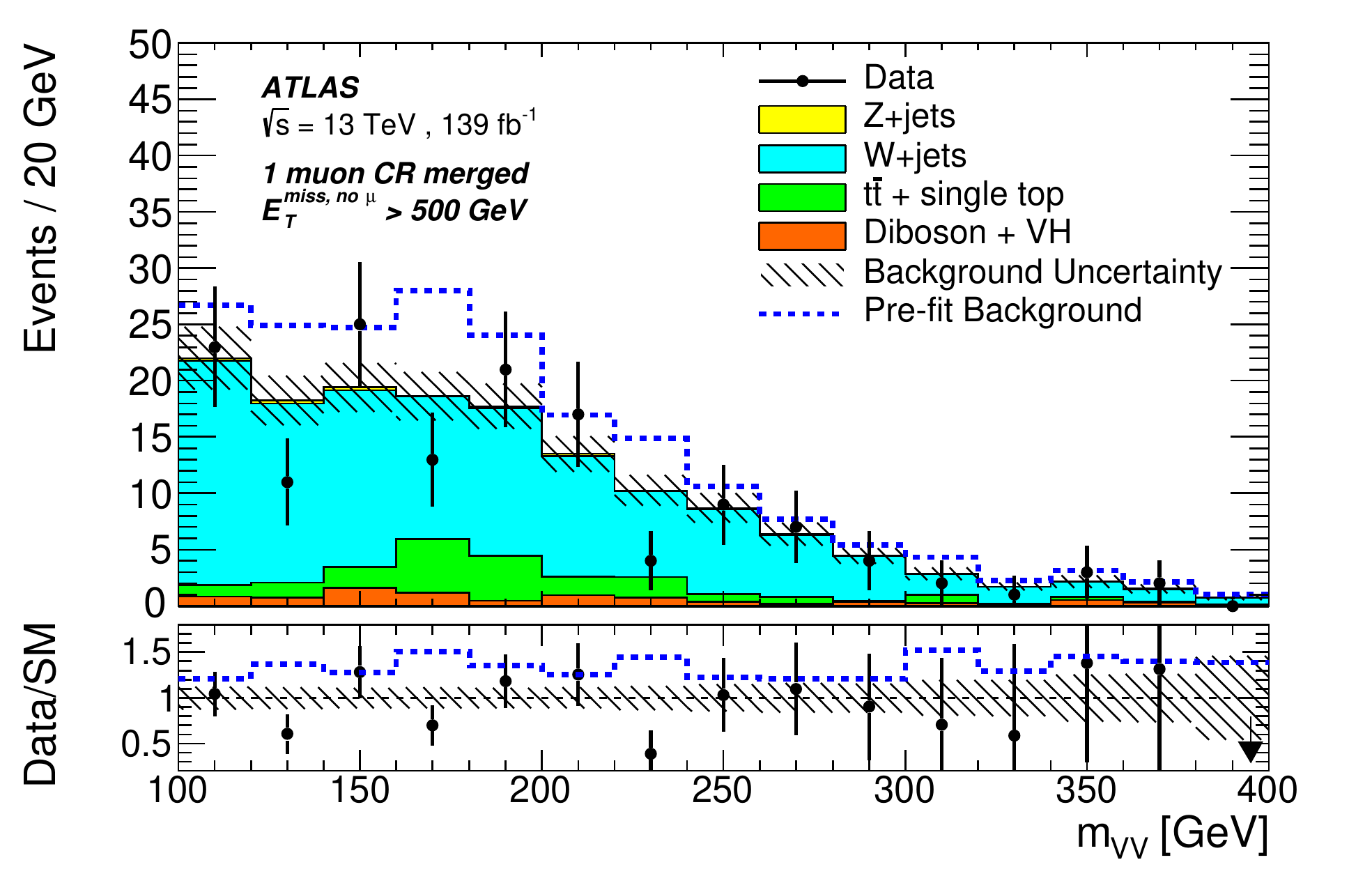}}
\subfloat[]{\includegraphics[width=0.49\textwidth]{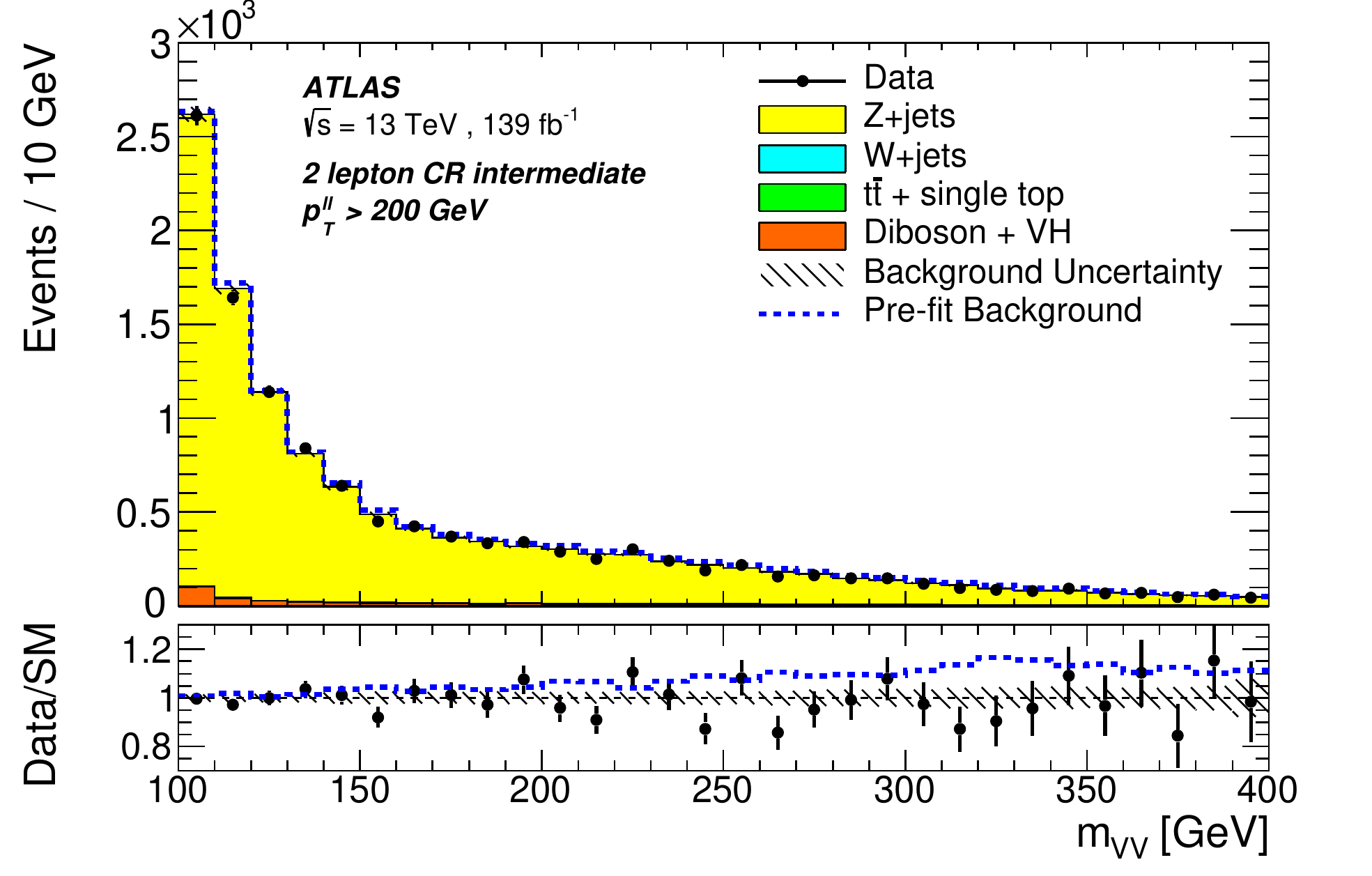}} \\
\subfloat[]{\includegraphics[width=0.49\textwidth]{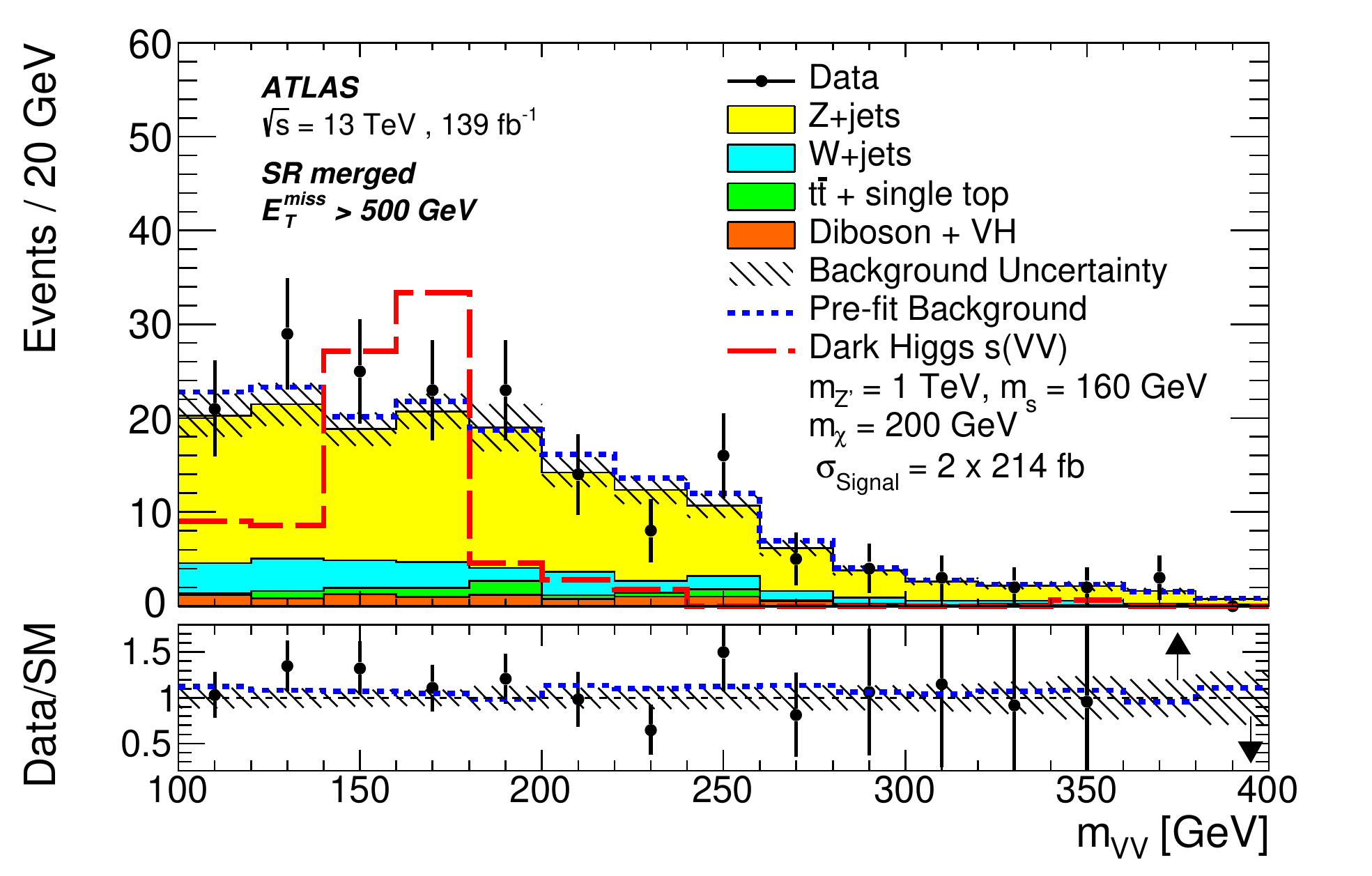}}
\subfloat[]{\includegraphics[width=0.49\textwidth]{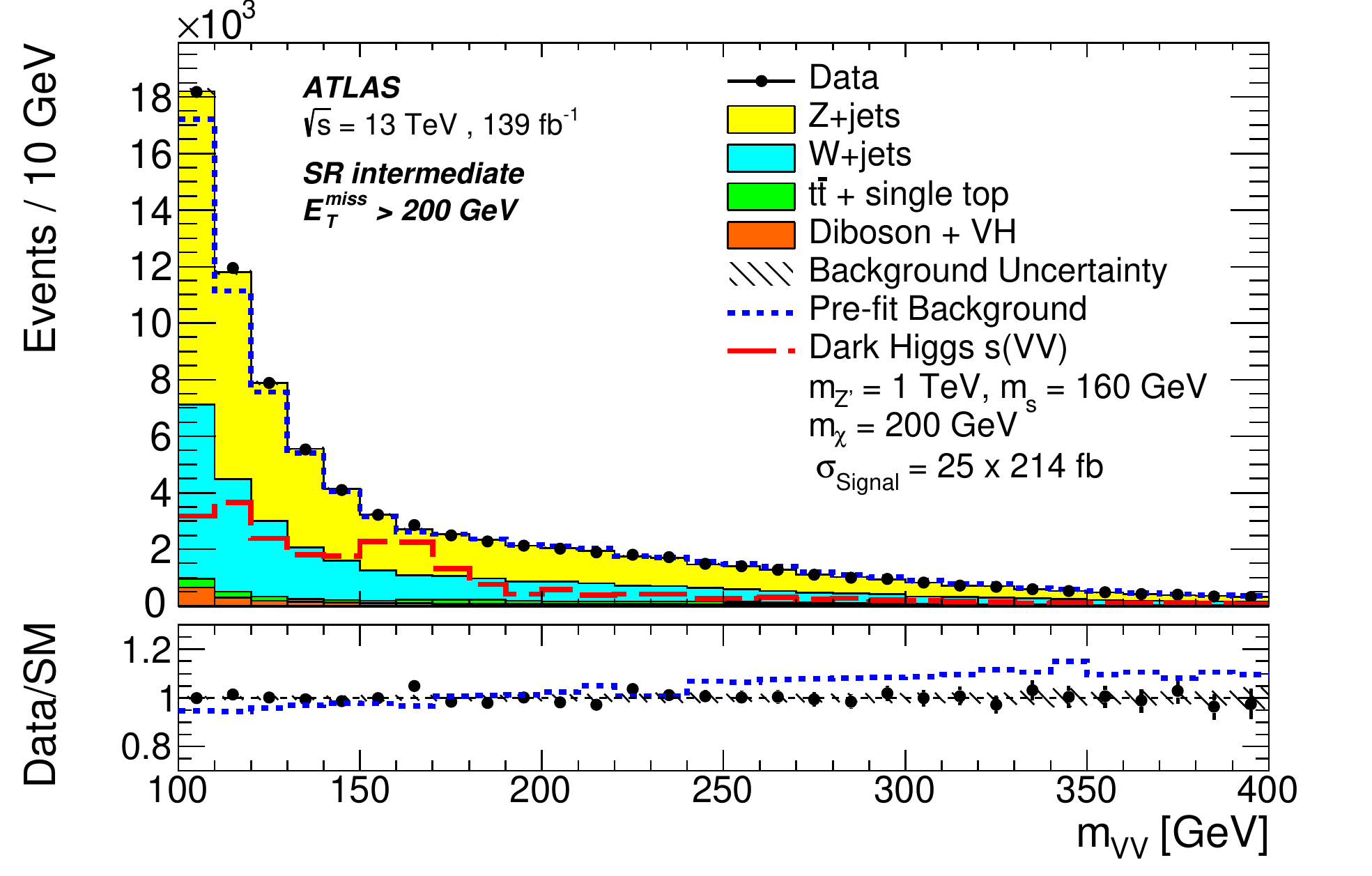}}
\caption{Distributions of the invariant mass of the dark Higgs boson candidates in the \onemucr and \twolepcr~(upper row) and in the SR~(lower row) in two representative categories, after the fit to data. The upper panels compare the data with the SM expectation before and after the background-only fit. The lower panels display the ratio of data to SM expectations after the fit, with its systematic uncertainty. Also shown is the ratio of SM expectations before and after the fit.
The expected signal, with a cross section of 214~fb, from a representative dark Higgs model with $\gq=0.25$, $\gchi=1.0$, and $\sin\theta=0.01$, is scaled for presentation purposes. No $m_{VV}$ shape information in the CRs is considered in the fit.
Pre-fit uncertainties cover differences between the data and pre-fit background prediction.
}
\label{fig:ms-body}
\end{figure}

Upper limits are set on the product of the $pp\rightarrow s \chi\chi$ production cross section and~$\mathcal{B}(\sVV)$, using a modified frequentist approach~(CL$_\mathrm{s}$)~\cite{Read:2002hq} with a test statistic based on the profile likelihood in the asymptotic approximation~\cite{Cowan:2010js}.
Exclusion contours in the (\mZp, \ms) plane for the dark Higgs model 
are presented in Figure~\ref{fig:limit_contour}, and exclude \mZp up to $1.8~\TeV$ for $\ms=210~\GeV$ at $95\%$ confidence level (C.L.). The observed exclusion range in \mZp becomes narrower than expected at low \ms owing to the small excess in data near $m_{VV}=160~\GeV$ discussed above.
The \merged SR provides the maximal sensitivity attained at low \ms and high \mZp, while the \intermediate SR provides complementary sensitivity.

\begin{figure}[h]
\centering
\includegraphics[width=0.6\textwidth]{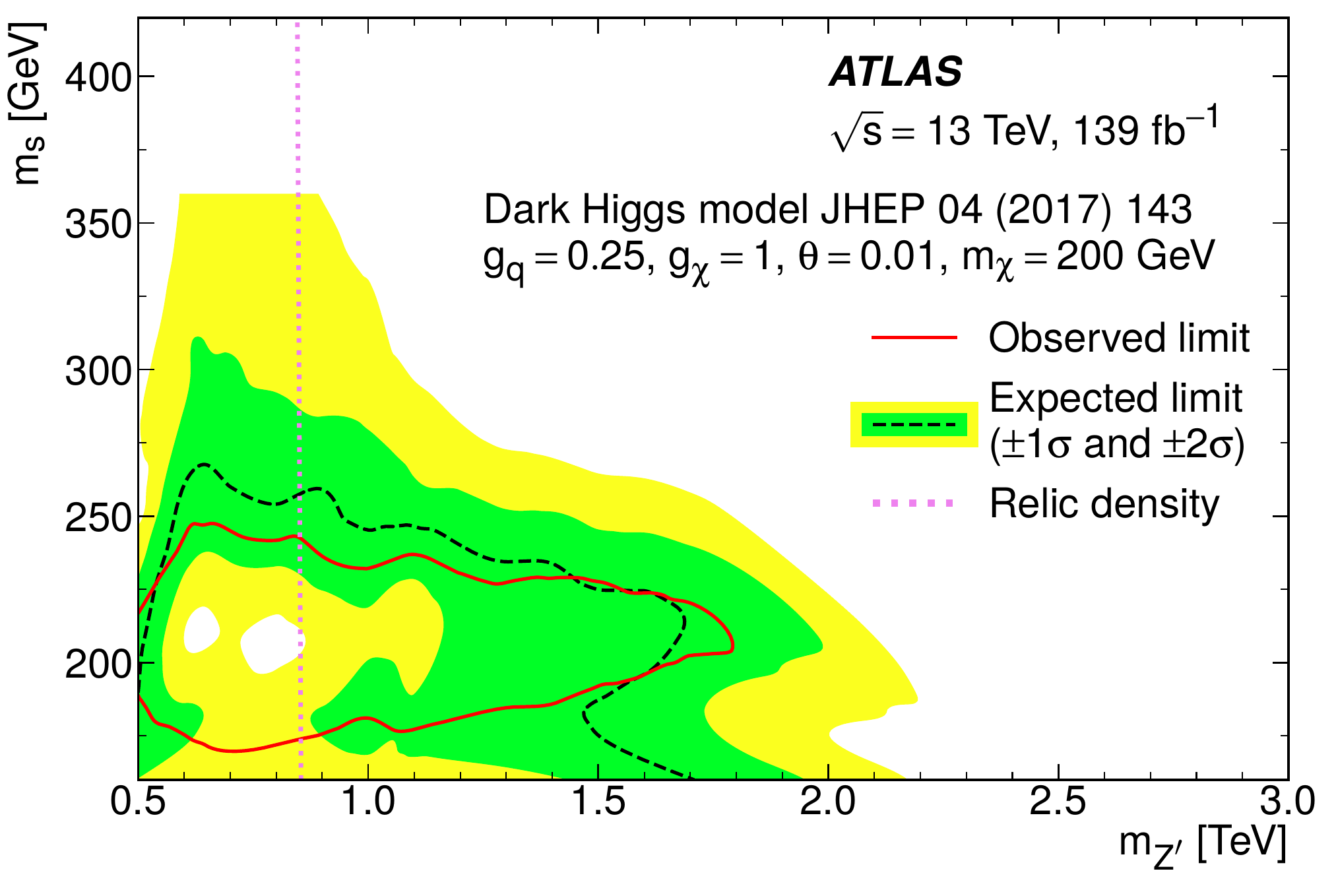}
\caption{
Observed (expected) exclusion regions at 95$\%$ C.L. for the dark Higgs model in the (\mZp, \ms) plane, encircled by the solid (dashed) line. The expected $\pm1\sigma$ ($\pm2\sigma$) uncertainty is shown as the filled green (yellow) band. The observed relic density~\cite{Aghanim:2018eyx} is obtained for $\mZp = 850~\GeV$ (dotted line).
}
\label{fig:limit_contour}
\end{figure}
 
\FloatBarrier

In conclusion, this Letter presents a novel search for DM in previously uncovered final states with large \met and hadronic decays of resonant $VV=W^\pm W^\mp$ or $ZZ$ pairs, with $m_{VV}>160~\GeV$, using the ATLAS detector at the LHC.
No significant excess over the predicted background is found in \Lumi~\ifb\ of 13~\TeV\ \pp collision data.
This search excludes previously uncharted parameter space of the dark Higgs model for $\ms>160~\GeV$, and provides sensitivity complementary to other DM searches using $X+\met$ signatures.

\section*{Acknowledgments}
 
% The next lines are included from the .//acknowledgements/Acknowledgements.tex input file

We thank CERN for the very successful operation of the LHC, as well as the
support staff from our institutions without whom ATLAS could not be
operated efficiently.
 
We acknowledge the support of ANPCyT, Argentina; YerPhI, Armenia; ARC, Australia; BMWFW and FWF, Austria; ANAS, Azerbaijan; SSTC, Belarus; CNPq and FAPESP, Brazil; NSERC, NRC and CFI, Canada; CERN; ANID, Chile; CAS, MOST and NSFC, China; COLCIENCIAS, Colombia; MSMT CR, MPO CR and VSC CR, Czech Republic; DNRF and DNSRC, Denmark; IN2P3-CNRS and CEA-DRF/IRFU, France; SRNSFG, Georgia; BMBF, HGF and MPG, Germany; GSRT, Greece; RGC and Hong Kong SAR, China; ISF and Benoziyo Center, Israel; INFN, Italy; MEXT and JSPS, Japan; CNRST, Morocco; NWO, Netherlands; RCN, Norway; MNiSW and NCN, Poland; FCT, Portugal; MNE/IFA, Romania; MES of Russia and NRC KI, Russia Federation; JINR; MESTD, Serbia; MSSR, Slovakia; ARRS and MIZ\v{S}, Slovenia; DST/NRF, South Africa; MICINN, Spain; SRC and Wallenberg Foundation, Sweden; SERI, SNSF and Cantons of Bern and Geneva, Switzerland; MOST, Taiwan; TAEK, Turkey; STFC, United Kingdom; DOE and NSF, United States of America. In addition, individual groups and members have received support from BCKDF, CANARIE, Compute Canada and CRC, Canada; ERC, ERDF, Horizon 2020, Marie Sk{\l}odowska-Curie Actions and COST, European Union; Investissements d'Avenir Labex, Investissements d'Avenir Idex and ANR, France; DFG and AvH Foundation, Germany; Herakleitos, Thales and Aristeia programmes co-financed by EU-ESF and the Greek NSRF, Greece; BSF-NSF and GIF, Israel; La Caixa Banking Foundation, CERCA Programme Generalitat de Catalunya and PROMETEO and GenT Programmes Generalitat Valenciana, Spain; G\"{o}ran Gustafssons Stiftelse, Sweden; The Royal Society and Leverhulme Trust, United Kingdom.
 
The crucial computing support from all WLCG partners is acknowledged gratefully, in particular from CERN, the ATLAS Tier-1 facilities at TRIUMF (Canada), NDGF (Denmark, Norway, Sweden), CC-IN2P3 (France), KIT/GridKA (Germany), INFN-CNAF (Italy), NL-T1 (Netherlands), PIC (Spain), ASGC (Taiwan), RAL (UK) and BNL (USA), the Tier-2 facilities worldwide and large non-WLCG resource providers. Major contributors of computing resources are listed in Ref.~\cite{ATL-SOFT-PUB-2020-001}.
 
% End of text imported from the .//acknowledgements/Acknowledgements.tex input file

\printbibliography
 
%\clearpage
%\part*{Auxiliary material}
%\addcontentsline{toc}{part}{Auxiliary material}
 
\clearpage
\input{atlas_authlist}\end{document}

%% file: atlas_authlist.tex
% ATLAS Collaboration author list
% Reference date of EXOT-2018-40 is 2020-07-06
% Author list last updated on date 13-OCT-20
% Data extracted on 13-Oct-2020 for paper reference EXOT-2018-40
% at 9:38am
 
\begin{flushleft}
\hypersetup{urlcolor=black}
{\Large The ATLAS Collaboration}

\bigskip

\AtlasOrcid[0000-0002-6665-4934]{G.~Aad}$^\textrm{\scriptsize 102}$,    
\AtlasOrcid[0000-0002-5888-2734]{B.~Abbott}$^\textrm{\scriptsize 128}$,    
\AtlasOrcid{D.C.~Abbott}$^\textrm{\scriptsize 103}$,    
\AtlasOrcid[0000-0002-2788-3822]{A.~Abed~Abud}$^\textrm{\scriptsize 36}$,    
\AtlasOrcid[0000-0002-1002-1652]{K.~Abeling}$^\textrm{\scriptsize 53}$,    
\AtlasOrcid[0000-0002-2987-4006]{D.K.~Abhayasinghe}$^\textrm{\scriptsize 94}$,    
\AtlasOrcid[0000-0002-8496-9294]{S.H.~Abidi}$^\textrm{\scriptsize 166}$,    
\AtlasOrcid[0000-0002-8279-9324]{O.S.~AbouZeid}$^\textrm{\scriptsize 40}$,    
\AtlasOrcid{N.L.~Abraham}$^\textrm{\scriptsize 155}$,    
\AtlasOrcid[0000-0001-5329-6640]{H.~Abramowicz}$^\textrm{\scriptsize 160}$
% ...
\bigskip
\\

% ...
$^{36}$CERN, Geneva; Switzerland.\\
% ...
$^{40}$Niels Bohr Institute, University of Copenhagen, Copenhagen; Denmark.\\
% ...
$^{53}$II. Physikalisches Institut, Georg-August-Universit\"{a}t G\"ottingen, G\"ottingen; Germany.\\
% ...
$^{94}$Department of Physics, Royal Holloway University of London, Egham; United Kingdom.\\
% ...
$^{102}$CPPM, Aix-Marseille Universit\'e, CNRS/IN2P3, Marseille; France.\\
$^{103}$Department of Physics, University of Massachusetts, Amherst MA; United States of America.\\
% ...
$^{128}$Homer L. Dodge Department of Physics and Astronomy, University of Oklahoma, Norman OK; United States of America.\\
% ...
$^{155}$Department of Physics and Astronomy, University of Sussex, Brighton; United Kingdom.\\
% ...
$^{160}$Raymond and Beverly Sackler School of Physics and Astronomy, Tel Aviv University, Tel Aviv; Israel.\\
% ...
$^{166}$Department of Physics, University of Toronto, Toronto ON; Canada.\\
% ...

\end{flushleft}

% Created with Glance <Atlas.Glance@cern.ch>